\documentclass{article}
\usepackage{amsfonts}
\usepackage{amssymb}
\newcommand{\nin}{\noindent}
\newcommand{\be}{\begin{equation}}
\newcommand{\ee}{\end{equation}}
\newcommand{\bea}{\begin{eqnarray}}
\newcommand{\eea}{\end{eqnarray}}

\newcommand{\nn}{\nonumber\\}

\newcommand{\ol}{\overline}

\newcommand{\den}{(q_0^2 -q^2 -q^4)}
\newcommand{\num}{(q_0\gamma_0 -q_k\gamma_k +q^2)}
\usepackage{graphicx}
\usepackage{caption}

\begin{document}

\begin{flushleft} 
KCL-PH-TH/2013-{\bf 20} 
\end{flushleft}

\vspace{1cm}

\begin{center}

{\Large{\bf Fermion effective dispersion relation for  $z=2$ Lifshitz QED}}

\vspace{0.5cm}

{\bf Jean Alexandre, James Brister}

\vspace{0.2cm}

King's College London, Department of Physics, WC2R 2LS, UK\\
jean.alexandre@kcl.ac.uk\\ 
james.brister@kcl.ac.uk

\vspace{1cm}

{\bf Abstract}

\end{center}

\vspace{0.5cm}

\nin We study consequences of Lorentz symmetry violation in a $z=2$ Lifshitz extension of QED in 3+1 dimensions, and we discuss non-trivial effects of
quantization. Because of the specific power of space momentum in propagators of the model, dimensional regularization leads 
to an unusual interpretation of loop integrals, which are finite even when the space dimension goes to the integer 3. We check the consistency 
of the approach by calculating the (vanishing) corrections to the photon mass and the IR-divergence-free corrections to the dispersion relation 
for massless fermions.

\vspace{0.5cm}

\section{Introduction}

Lifshitz models, based on an anisotropy between space and time, and therefore breaking Lorentz symmetry, have been motivated by the 
possibility of defining new renormalizable interactions. These models involve a mass scale $M$, typically of the order of a Grand Unified Theory (GUT) scale,
bellow which the classical dispersion relations are almost relativistic, with corrections suppressed by powers of the ratio  
$k/M$, where $k$ is the space momentum of the particle (see \cite{Alex} for a review). 
In the infrared (IR), at the classical level, these deviations from a relativistic dispersion relation can easily be 
within the experimental bounds of Lorentz symmetry violation \cite{KosteleskyRussel}, but taking into account quantum corrections changes this picture.
Due to the above mentioned anisotropy, space and time derivatives are not dressed in the same way by quantum corrections. 
If one considers only one kind of particle, a rescaling of space coordinates then allows one to recover a relativistic dispersion relation in the IR, after 
quantum corrections. But in a more realistic situation, with several species interacting, rescaling fields and coordinates cannot lead
to a relativistic IR dispersion relation for all particles.
This was first noted in \cite{IengoRussoSerone}, where the unnatural fine tuning of bare parameters is shown, in order to  match the light cones seen by
two different scalar particles interacting. A similar study was done in \cite{AlexBristerHouston}, where the effective dispersion relation for 
interacting Lifshitz fermions is derived, in the case where flavour symmetry is broken. It was shown that the 4-fermion interaction, renormalizable in 
the Lifshitz context, generates corrections to the relativistic IR dispersion relation, which diverge quadratically with the cut off, therefore limiting
the phenomenological viability of the model.

We note that in \cite{mdynLifshitz} and \cite{AlexBristerHouston}, dynamical mass generation was also studied, 
which is another essential feature of Lifshitz models, besides the introduction of new renormalizable interactions, 
and which is possible because Lorentz symmetry violation introduces a mass scale in the problem.
This is similar to the occurrence of dynamical mass generation for  Quantum Electrodynamics (QED) in the presence of an external 
magnetic field \cite{mag}, which provides an effective Lorentz symmetry violation.

Our aim is to study here the phenomenological viability of a Lifshitz QED model. General properties of 
Lifshitz gauge theories have been studied in \cite{Anselmi}, and we focus here on one specific example, with an anisotropy scaling $z=2$,
by calculating the one-loop correction to the effective ``maximum'' speed $v$ seen by fermions. 
A similar study has been done in \cite{AlexVergou}, for a Lorentz-symmetry-violating extension of QED - not of a Lifshitz type though - 
where a rescaling of fields and spacetime coordinates, after one-loop corrections, leads to a usual relativistic IR dispersion relation for the photon 
but to a subluminal propagation for fermions.

$z=2$ Lifshitz QED is super-renormalizable, but still contains power counting diverging graphs. Nevertheless, the would-be divergent 
graphs we calculate here happen to be finite in $3-\epsilon$ space dimensions, after integration over frequencies, even in the limit $\epsilon\to0$.
This special feature is a consequence of the specific powers of the momentum in this model. Dimensional regularization is still
needed though, in order to define the would-be ultraviolet (UV) diverging integrals. But because we have $z=2$, the limit $\epsilon\to0$ never leads to 
a pole of the Gamma function and the graphs are UV finite. The consistency of the approach is checked with the vanishing of the photon mass correction.
Divergences in scalar Lifshitz models are discussed for different values of $z$ and of space dimension in \cite{effpot}, where, in the case of scalar QED, 
the effective potential for the scalar field is finite for $z=2$ and in 3 space dimensions.

Another feature related to dimensional regularization of this model is the appearance of poles in $\epsilon$, but due to IR divergences in 
the case of massless fermions. In order to calculate the integrals analytically, we will consider this massless limit, since we are interested in the effective
fermion dispersion relation, which is independent of the mass anyway. As expected, the poles in $\epsilon$ generated by IR divergence cancel
each other in the calculation of the effective ``maximum'' speed $v$.

Section 2 presents the model and its properties, from which the one-loop self energies and fermion dispersion relation
are calculated in section 3, with details in the Appendix.
We eventually find that the model predicts a too important deviation from IR relativistic kinematics at one-loop, 
if one identifies the dimensionless coupling with the fine structure constant,
whereas the classical theory is totally justifiable in the IR regime.

\section{Model}

We consider here the following Lagrangian density for a $z=2$ Lifshitz QED model, with metric (1,-1,-1,-1),
\be
{\cal{L}} = \frac{1}{2}F_{0i}F_{0i} - \frac{1}{4}F_{ij}(M^2 -\Delta)F_{ij} + \ol{{\psi}}(iD_0 \gamma_0 - iM D_k \gamma_k -D_k D_k -m^2)\psi
\ee
Where $D_\mu = \partial_\mu +ieA_\mu$ and $F_{\mu\nu} = \partial_{[\mu}A_{\nu]}$.
The mass dimensions are, in 3+1 dimensions,
\be
[A_j]=\frac{1}{2}~~, ~~ [A_0]=[\psi]=\frac{3}{2}~~,~~[e]=\frac{1}{2}~~,~~[M]=[m]=1~.
\ee
The theory is invariant under the $U(1)$ gauge symmetry
\be
\psi \rightarrow \psi e^{i\Lambda}~~,~~
\ol{\psi}  \rightarrow \ol{\psi}  e^{-i\Lambda}~~,~~
A_\mu  \rightarrow A_\mu +\frac{1}{e} \partial_\mu \Lambda~,
\ee
and the classical dispersion relations are, for the photon and the fermion respectively,
\bea
k_0^2&=&M^2k^2+k^4\nn
k_0^2&=&M^2k^2+(k^2+m^2)^2~,
\eea
where $k_0$ denotes the frequency and $k=\sqrt{k_ik_i}$ the space momentum. 
After the rescaling $k_0\to M k_0$, the dispersion relations become
\bea\label{classicaldisprel}
k_0^2&=&k^2+\frac{k^4}{M^2}~~~~~~\mbox{(photon)}\nn
k_0^2&=&m_R^2+(1+\eta)k^2+\frac{k^4}{M^2}~~~~~\mbox{(fermions)},
\eea
where the rescaled mass is $m_R\equiv m^2/M$ and $\eta\equiv2m^2/M^2$. \\
To get a idea of the orders of magnitude, let us consider the electron 
mass $m_R\simeq0.5$ MeV and the GUT scale $M\simeq10^{16}$ GeV. We have then $m^2\simeq5\times10^{12}$ GeV$^2$ and $\eta\simeq10^{-19}$, 
which is well within the Lorentz symmetry violation bounds \cite{KosteleskyRussel}. Also, the corrections 
$k^4/M^2$ are by far currently not detectable, for energies at most $k\sim10$ TeV.
For this reason, classical Lifshitz models can be realistic phenomenologically. But, as we will stress in this article, quantum corrections
completely change this picture.

\subsection{Propagators and interactions}

The fermion propagator is given by
\be
G(p_0,p) = i\frac{p_0 \gamma_0 -  M p_i \gamma_i + p^2 + m^2}{p_0^2 -M^2 p^2 - (m^2 +p^2)^2}
\ee
In order to define the photon propagator, 
we impose the Feynman-like condition 
\be\partial_0 A_0 -(-\Delta+ M^2)\partial_k A_k =0~, 
\ee
which leads to a non-local gauge-fixing term in the Lagrangian \cite{Anselmi2}, but to a well-behaved photon propagators.
The gauge fixing term is
\be
{\cal{L}}_{GF} = -(\partial_0 A_0 -(-\Delta+ M^2)\partial_i A_i) \frac{1}{2(-\Delta+ M^2)} (\partial_0 A_0 -(-\Delta + M^2)\partial_j A_j)
\ee
Throughout this paper, we shall treat the $A_0$ and $A_i$ lines in graphs separately, as they give different contributions, and
the photon propagator is then
\be
D_{00}(k_0,k) = -i\frac{M^2+ k^2}{k_0^2 -M^2 k^2 -k^4}
\ee
\be
D_{ij}(k_0, k) = i\frac{\delta_{ij}}{k_0^2 -M^2 k^2 -k^4}
\ee
with no off-diagonal components (before quantum corrections).\\
There are three  vertices: the 3-point $\ol{\psi}A_i \psi$, contributing $-ie(\gamma_i M +2p^{(\psi)}_i +p^{(A)}_i)$; 
the 3-point $\ol{\psi}A_0 \psi$, contributing $ie\gamma_0$ and the 4-point $A_iA_j\psi \ol{\psi}$, contributing $-2ie^2 \delta_{ij}$.\\
From the expressions for the vertices and propagators given above, the superficial divergence for an arbitrary graph is found to be
\be
D = 5L -2I_{A_0} -4I_{A_i} -2I_\psi + N_{3i}~,
\ee 
where $I_x$ is the number of internal lines of particle $x$, $L$ the number of loops and $N_{3i}$ the number of three-point 
spacial-photon vertices. With the standard relations 
$L = 1 + \Sigma_xI_x - \Sigma_a N_a $ and $(2I_x + E_x) = \Sigma_a n_a N_a$ (for each species $x$, where $n_a$ is the number 
of $x$-lines on the vertex $a$) we can re-write the above in terms of external lines, $E$, as
\be\label{D}
D = 6 - E_{A_i} -2E_{A_0} -2E_\psi -L~.
\ee
In this article, after quantum corrections we will rescale fields and coordinates such that the photon has the usual IR relativistic
dispersion relation, and we will look at the consequence on the fermion dispersion relation. From eq.(\ref{D}), the superficial degree of
divergence of the fermion propagator is 1 at one loop, such that the divergence of the fermion wave function renormalization is logarithmic.

\subsection{Symmetries of the dressed propagators}

As in usual QED, the polarisation tensor $\Pi_{\mu\nu}$ is transverse, since this property arises from the structure of 
the source terms in the partition function,
and is independent of the details of the kinetic term for the gauge field. The most general form for $\Pi_{ij}$ is, at quadratic order 
in  the momentum,
\be
\Pi_{ij}=ZM^2(k_i k_j - k^2 \delta_{ij})+Wk_0^2\delta_{ij}~,
\ee
where $Z,W$ are dimensionless corrections to be calculated.
Transversality $k^\mu\Pi_{\mu\nu}=0$ implies that then that the other components of the polarization tensor are of the form
\be
\Pi_{0i}=Wk_0k_i~~~,~~~\Pi_{00}=Wk^2~,
\ee
such that the photon dressed propagator ${\cal D}_{\mu\nu}$ satisfies
\bea \label{Ph}
{\cal D}_{00}^{-1}(k_0,k) &=& D_{00}^{-1}(k_0,k) + W k^2 \nn
{\cal D}_{ij}^{-1}(k_0,k) &=& D_{ij}^{-1}(k_0,k) + ZM^2(k_i k_j - k^2 \delta_{ij}) +Wk_0^2\delta_{ij} \nn
{\cal D}_{0i}^{-1}(k_0,k) &=& W k_0k_i~. \nonumber
\eea
The fermion dressed propagator is denoted, at the lowest order in momentum,
\be \label{Fe}
{\cal G}^{-1}(p_0,p) = G^{-1}(p_0,p) +\delta m^2 + Xp_0 \gamma_0 - Yp_i \gamma_i M ~,
\ee
where $X,Y$ are dimensionless corrections to be calculated.

\section{One-loop propagators}

\begin{figure}\label{graphs}
	\includegraphics [width=0.48\textwidth ]{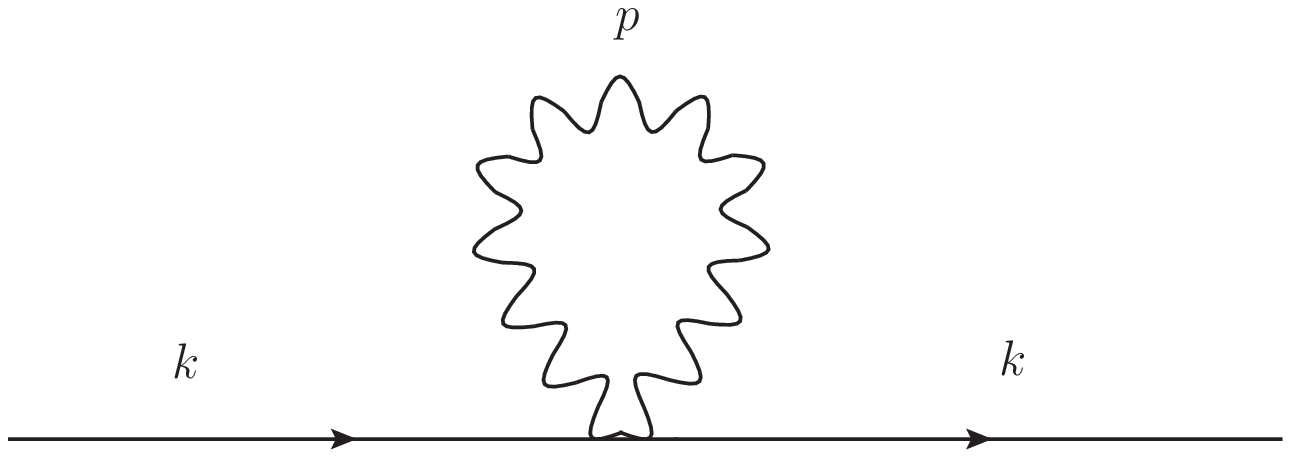}
	\includegraphics [width=0.48\textwidth ]{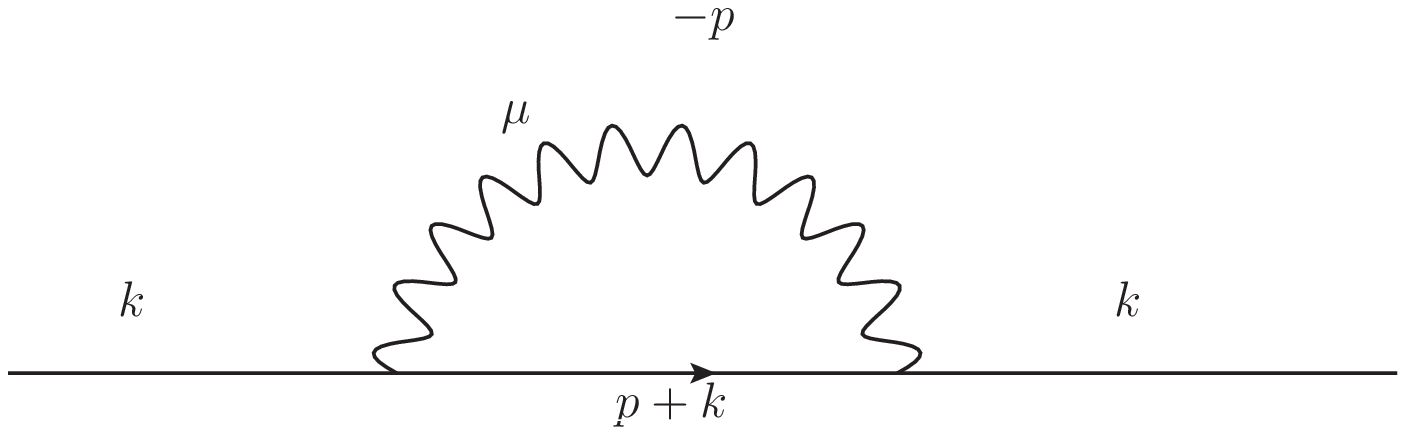}
	\includegraphics [width=0.48\textwidth] {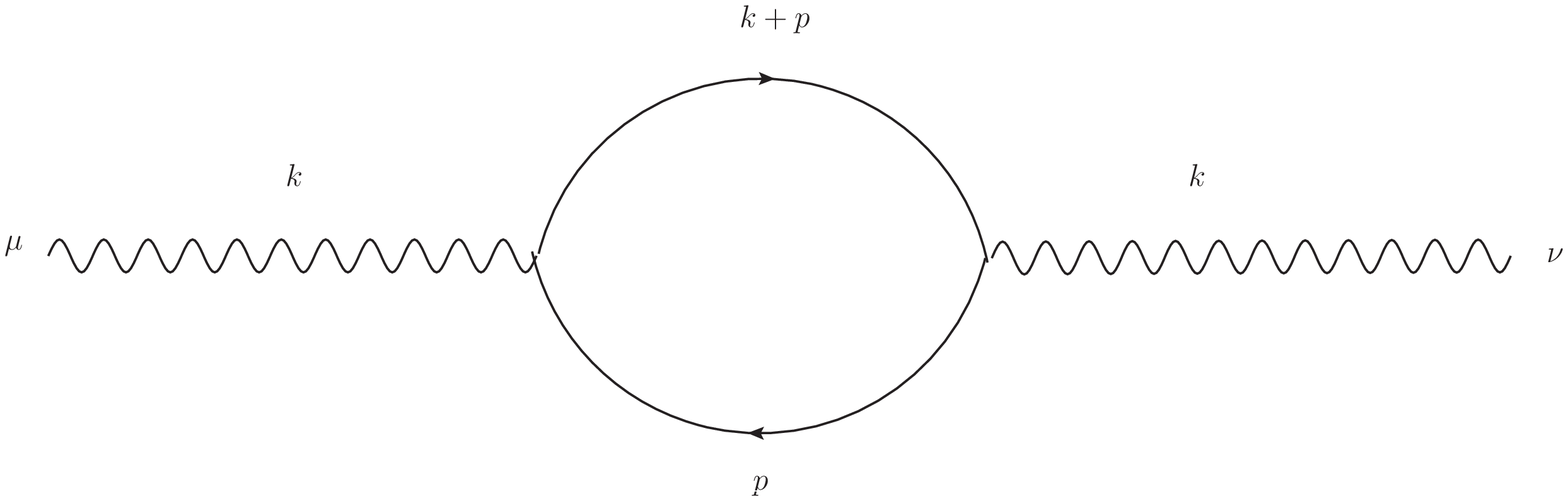}
	\includegraphics [width=0.48\textwidth] {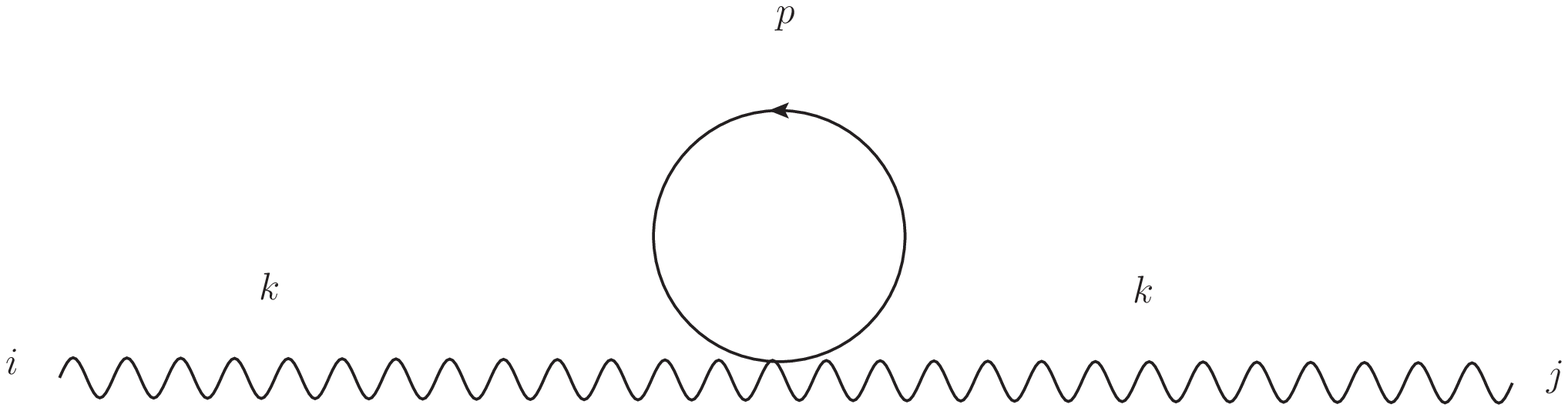}
	\caption{Relevant graphs for the evaluation of photon and fermion self energies.}
\end{figure}

The one-loop graphs contributing to the two-point functions are shown in fig.1; for both fermions and photons, 
the graphs involving the four-point function do not depend on the external momentum, $k$, and so do not contribute 
to the wave function renormalisation.

\subsection{IR behaviour}

In the following sections we shall take the fermion mass to be zero, as this greatly simplifies the calculations.
This will unfortunately introduce IR divergences, which are regularised by dimensional continuation to $3-\epsilon$ spatial dimensions. 
Nevertheless, as we shall show, these divergences cancel out in the calculation of corrections to the dispersion relations.

\subsection{UV behaviour}

From naive power counting (see eq.(\ref{D})), one would expect that the terms $Y$ and $Z$ would be divergent, 
logarithmically and linearly, respectively. However, dimensional regularization leads to integrals of the form 
\be
\int_0^\infty \frac{q^{2a}}{(q^2+q^4)^{r-\frac{1}{2}}} dq~,
\ee
(for some integers $a,r$), which, after dimensional continuation $2a \rightarrow 2a-\epsilon$ gives, for appropriate values of $\epsilon$
\be
\frac{\Gamma \left(a-r-\frac{\epsilon }{2}+1\right) \Gamma 
\left(- a+2 r+\frac{\epsilon}{2} -\frac{3}{2})\right)}{2 \Gamma \left(r-\frac{1}{2}\right)}
\ee
which can never lead to a pole of the second Gamma function in the numerator in the limit $\epsilon \rightarrow 0$, usually responsible 
for UV divergences. The first Gamma function may be divergent for sufficiently high $r$ or low $a$, but this is an IR divergence.
To check the consistency of this unusual feature, we show in the Appendix \ref{photonmass} that corrections to the photon mass indeed vanish, as expected 
from gauge invariance.

\subsection{One-loop corrections}

Details of the calculations for the self energies can be found in the Appendix, and we find
\bea\label{XYWZ}
X&=& \frac{-e^2}{8\pi^2M} \left( \frac{1}{\epsilon} + \frac{3}{2} - \frac{A}{2} \right) +O(\epsilon) \\
Y&=& \frac{-e^2}{8\pi^2M} \left( \frac{1}{\epsilon} + \frac{7}{2} - \frac{A}{2} \right) +O(\epsilon) \nn
W&=& \frac{-e^2}{6\pi^2M} \left( \frac{-1}{\epsilon} -\frac{5}{3} + \frac{A}{2} \right) +O(\epsilon)\nn
Z &=& \frac{-e^2}{6\pi^2M} \left( \frac{-1}{\epsilon} +2 + \frac{A}{2} \right) +O(\epsilon) ~,\nonumber
\eea
where $A= \gamma_E +2 \log 2 - \log \pi$ and we remind that the poles in $\epsilon$ correspond to IR divergences due to the massless fermion limit.\\
With the corrections calculated above, the IR kinetic terms of the one-loop dressed Lagrangian are
\bea
{\cal{L}}_{IR} &=& \frac{1}{2}(1+W)F_{0i}F_{0i} - \frac{1}{4}(1+Z)F_{ij}F_{ij}M^2\\ 
&&+ \ol{{\psi}}(i\partial_0 \gamma_0(1+X)-iM \partial_k \gamma_k(1+Y))\psi~, \nonumber
\eea
and, in order to recover the relativistic IR propagation for photons, we rescale
\bea\label{rescale}
k_0 &\rightarrow& M\frac{(1+Z)^{1/4}}{(1+W)^{1/2}}~k_0\\
k_i &\rightarrow& \frac{1}{(1+Z)^{1/4}}~k_i\nn
A_0 &\rightarrow& M^{1/2}\frac{(1+Z)^{1/4}}{(1+W)^{1/2}}~A_0\nn
A_i &\rightarrow& \frac{1}{M^{1/2}(1+Z)^{1/4}}~A_i\nn
\psi&\rightarrow&\frac{(1+W)^{1/4}}{(1+X)^{1/2}(1+Z)^{1/8}}~\psi~.\nonumber
\eea
This leads to the following effective kinetic Lagrangian
\be\label{effective}
M^{-1}{\cal L}_{IR}^{eff}=-\frac{1}{4}F_{\mu\nu}F^{\mu\nu}+i\ol\psi\left[\gamma^0\partial_0
-\left(1+\frac{\delta v}{2}\right)\vec\gamma\cdot\vec\partial\right]\psi~,
\ee 
where
\be
\delta v\equiv 2Y-2X+W-Z~,
\ee
and the factor $M^{-1}$ is absorbed by the time rescaling $t\to Mt$ in the definition of the action.
As expected, the photon dispersion relation is relativistic in the IR
\be
k_0^2 = k^2~~~~~~~\mbox{(photon)}~,
\ee
but the one-loop dispersion relation for fermions becomes
\be
k_0^2 = m_R^2 + (1+\delta v)k^2~~~~~~\mbox{(fermions)}~,
\ee
where $m_R$ is the rescaled and dressed one-loop fermion mass, which we haven't taken into account up to now
\footnote{Note that, since we did the calculations in the massless fermion case, the classical modification $1\to(1+\eta)$ in eq.(\ref{classicaldisprel}) 
is not present here. But, as explained below eq.(\ref{classicaldisprel}), this contribution is negligible compared to
the present correction due to quantum effects: $\eta<<\delta v$.}.
The product of phase and group velocities for fermions is $v_pv_g=1+\delta v$, and 
from the expressions (\ref{XYWZ}), we can see that the $1/\epsilon$ terms cancel, leading to
\bea
\delta v = \frac{1}{9\pi^2}\frac{e^2}{M}~.
\eea
Note that the superluminal propagation is a natural feature of Lifshitz models.
We stress again that the rescaling laws (\ref{rescale}) are UV finite, and that the poles in $\epsilon$ correspond to ``artificial'' IR divergences,
as a consequence of the massless fermion case we study here. The expected cancellation of these artificial poles in $\epsilon$ shows the consistency of the 
definition of loop integrals, which should lead to an IR-divergence-free expression for $\delta v$.

\subsection{Phenomenology}

If we identify the dimensionless coupling $e^2/M$ with $4\pi \alpha$, where $\alpha\simeq1/137$ is the fine-structure constant, 
we obtain 
\be
\delta v\simeq10^{-3} ~,
\ee
which is phenomenologically not realistic. The corresponding Lorentz-violating operators in the Standard Model Extension (SME) \cite{CK} are
parametrized by the CPT-even coefficients $c_{\mu\nu}$, defined as
\be
ic_{\mu\nu}\ol\psi\gamma^\mu \partial^\nu\psi~.
\ee
In our case, we have $2c_{ij}=\delta v~\delta _{ij}$, such that $\delta v=(2/3)\mbox{tr}\{c_{ij}\}$. From the tables of SME coefficients \cite{KosteleskyRussel},
the later identification gives an upper bound of the order $|\delta v|\lesssim 10^{-15}$, such that
one needs to consider a dimensionless coupling which satisfies
\be
\frac{e^2}{4\pi M}\lesssim 10^{-14} <<\alpha~.
\ee

We conclude this section with a remark concerning the dressed coupling constant of the model.
From the effective Lagrangian (\ref{effective}), gauge invariance ensures that the IR effective interactions are
\be
-\frac{e}{M^{1/2}}\ol\psi\left[\gamma^0 A_0-\left(1+\frac{\delta v}{2}\right)\vec\gamma\cdot\vec A\right]\psi~,
\ee
such that one can define the one-loop effective coupling $e^{(1)}$ as 
\be
e^{(1)}\equiv e\left(1+\frac{\delta v}{2}\right)~.
\ee
Although $e^{(1)}$ does not contain UV divergence, one can define the one-loop beta function for the evolution of the dressed coupling 
$e^{(1)}$ with the scale $M$, for fixed bare coupling $e$, which leads to 
\be\label{beta}
\beta\equiv M\frac{\partial e^{(1)}}{\partial M}=-\frac{1}{18\pi^2}\frac{e^3}{M}~.
\ee
Note that this definition of beta function does not coincide with the usual one, but rather shows how the effective coupling of the
model evolves as the crossover scale $M$ between Lifshitz and relativistic regimes varies.

\section{Conclusion}

We showed, with a specific Lifshitz extension of QED which is classically acceptable from the phenomenological point of view, that 
quantum corrections change the naive picture and lead to non-trivial  Lorentz violating effects. In order to recover a realistic
theory, the model must satisfy strong constraints.

Technically speaking, it is interesting to see that, because of higher order space derivatives, the unusual powers of space momentum 
lead to specific regularization features. Dimensional regularization leads to a definition of loop integrals which makes sense even in the 
limit where the space dimension goes to 3. The approach is nevertheless consistent, since the physical quantities calculated here
show the appropriate cancellations of UV would-be divergences (see calculation of quantum corrections for 
the photon mass) and cancellation of IR divergences in the case where fermions are massless (see calculation of fermion dispersion relation).

We also note that taking the limit $M\to\infty$ of the present model is not straightforward. Indeed, since the relevant (dimensionless) coupling is $e^2/M$,
this limit cannot be taken for fixed bare parameter $e$ without the theory becoming trivial. For this reason, the specific features of the 
present model do not allow one to continuously recover usual QED in the limit $M\to\infty$.

\vspace{1cm}

\nin{\bf Acknowledgements} J. A. would like to thank Kostas Farakos for useful suggestions.

\appendix

\section{Details of loop integrals}

Throughout this section, we shall use dimensionless momentum 4-vectors. In the following, $k_\mu = (M^2 r_0, M \vec r)$ 
is the external momentum, and $p_\mu = (M^2 q_0 , M \vec q)$ is a loop momentum to be integrated over (for scalar $q_0$ and 3-vector $q$).
The graphs are calculated by first integrating over frequencies and then using the three-dimensional 
spherical co-ordinates, with external space momentum $\vec r= (0,0,r)$ and loop momentum $\vec q=(q \cos \theta \sin \phi,q \sin \theta \sin \phi,q \cos \phi)$. 
In all but two relevant cases (those used to calculate $Z$ and the space part of the fermion integral) the integrands are a function of only $|q|, q_0$ 
and the angle $\phi$ between $\vec r$ and $\vec q$. 
The integration measure for $q$ in $3-\epsilon$ dimensions can therefore be written as:
\be
\int d^{3-\epsilon}q = \int_0^\infty q^{2-\epsilon} dq \int_0^\pi \sin^{1-\epsilon}\phi d \phi \int d\Omega_{2-\epsilon}
\ee
Where $d\Omega_{2-\epsilon}$ represents the integral over the remaining angular variables, which evaluates to 
\be
2\frac{ \pi^{\frac{2-\epsilon}{2}}}{\Gamma(\frac{2-\epsilon}{2})}
\ee
In the few cases where the remaining angles cannot be eliminated, the index structure shows that they will appear only at quadratic order, 
as $q_1^2$ or $q_2^2$. We can thus safely make the substitution $q_1^2 \rightarrow q^2 \frac{1}{2-\epsilon} \sin^2\phi $ wherever such terms appear. 

In what follows, we note
\be
\int_q\equiv\int \frac{d^{3-\epsilon} q dq_0 }{(2\pi)^{4-\epsilon}}~.
\ee

\subsection{Photon mass}\label{photonmass}

As a check, we shall ensure the photon mass correction vanishes, as implied by the Ward identity. For ${\cal D}_{00}$ and ${\cal D}_{0i}$ 
we need consider only the vacuum polarisation graph, but for ${\cal D}_{ij}$ we must consider both photon graphs shown in the two lower
figures of fig. 1. \\
The $(i,0)$ term is proportional to the integral
\be
\mbox{Tr}\int_q\frac{(\gamma_i +2q_i)\num \gamma_0 \num}{\den^2}~.
\ee
Every term in the numerator is either traceless or proportional to an odd power of $q$ and thus vanishes.\\
The $(0,0)$ term is
\be
\mbox{Tr}\int_q \frac{\gamma_0\num \gamma_0 \num}{\den^2} 
= \int_q \frac{4(q_0^2+q^2+q^4)}{\den^2}=0~.
\ee
This vanishes due to the $q_0$ integration and so is unaffected by dimensional regularisation. \\
Finally, the $(i,j)$ term from the vacuum polarisation is
\be
\mbox{Tr}\int_q \frac{(\gamma_i +2q_i)\num(\gamma_j +2q_j) \num}{\den^2}~,
\ee
which, with dimensional regularisation, gives a finite value of $-4/3\pi^2$ in the limit $\epsilon \rightarrow 0$ 
(which we can take as there is no IR divergence). Similarly, the graph with a four-point vertex gives
\be
2\mbox{Tr}\int_q \frac{\num}{\den}~,
\ee
which evaluates to $4/3\pi^2$, canceling the other contribution.

\subsection{Fermion corrections}

Writing $e \Gamma_\mu$ for the vertex functions, with $\Gamma_0 = \gamma_0, \Gamma_i = M\gamma_i + 2p^{\psi}_i + p^{A}_i$,   
we are looking for the terms linear in $r_0$ and $r$ in the expression
\bea
&&e^2 M^{1-\epsilon} \int_q~ \Gamma^\mu ~ G(q+r)~\Gamma^\nu D_{\mu\nu}(q) \\
&=& e^2  M^{1-\epsilon}  \int_q\bigg\{
-\frac{\gamma_0[\gamma_0(q_0+ r_0) -\gamma_l(q_l +r_l)+ (q+r)^2]\gamma_0(q^2+1)}{\den ((q_0+r_0)^2 -(q+r)^2 (q+r)^4)}\nn
&+&\frac{(\gamma_i+2r_i+q_i)[\gamma_0(q_0+ r_0) 
-\gamma_l(q_l +r_l)+ (q+r)^2](\gamma_i+2r_i+q_i)}{\den ((q_0+r_0)^2 -(q+r)^2 (q+r)^4)} \bigg\}~,\nonumber
\eea
where $G$ and $D$ are the fermion and photon propagators, respectively. Using the notation of eq.(\ref{Fe}), we find
\bea
Y &=& \frac{-e^2}{M} \bigg( (3-\epsilon)\frac{2^{\epsilon-7}\pi^{-2+\frac{\epsilon}{2}}(3+\epsilon-\epsilon^2)\Gamma(\frac{-\epsilon}{2}) 
\Gamma(\frac{1+\epsilon}{2})}{\Gamma(\frac{5-\epsilon}{2})} \nn 
&&~~~~~~~~~~~~~~-\frac{2^{\epsilon-6}\pi^{-2+\frac{\epsilon}{2}}\Gamma(\frac{-\epsilon}{2}) 
\Gamma(\frac{1+\epsilon}{2})}{\Gamma(\frac{3-\epsilon}{2})}  \bigg)\nonumber \\
&=& \frac{-e^2}{8M\pi^2} \left( \frac{1}{\epsilon} + \frac{7}{2} - \frac{A}{2} \right) +O(\epsilon)~,
\eea
where $A= \gamma_E +2 \log 2 - \log \pi$. Similarly, we find
\be
X= \frac{-e^2}{8\pi^2M} \left( \frac{1}{\epsilon} + \frac{3}{2} - \frac{A}{2} \right) +O(\epsilon)~,
\ee
so that the contribution to the velocity correction is
\be
2(Y-X)=  \frac{-e^2}{2\pi^2M}~.
\ee

\subsection{Photon corrections}

We shall calculate $W$ from the term quadratic in external momentum of ${\cal D}_{00}$,  hence we need the terms proportional to $r^2$ from the integral
\be
e^2  M^{1-\epsilon}  ~\mbox{Tr}
\int_q \frac{\gamma_0(\gamma_0(q_0+ r_0) -\gamma_l(q_l +r_l)+ (q+r)^2)\gamma_0\num }{\den ((q_0+r_0)^2 -(q+r)^2 -(q+r)^4)}~,
\ee
and we find
\bea
W &=&\frac{-e^2}{M}  \frac{2^{\epsilon-3}\pi^{-2+\frac{\epsilon}{2}}(2\epsilon-3)\Gamma(\frac{-\epsilon}{2}) 
\Gamma(\frac{3+\epsilon}{2})}{3\Gamma(\frac{5-\epsilon}{2})} \nonumber \\
&=& \frac{-e^2}{6\pi^2M} \left( \frac{-1}{\epsilon} -\frac{5}{3} + \frac{A}{2} \right) +O(\epsilon)~.
\eea
To calculate the term $Z$, we need to evaluate ${\cal D}_{ij}$ at one loop. Having done so in $d=3$, 
we obtain a matrix $\Pi_{ij}$ of integrals, being the terms proportional to $r^2$ in 
\bea
&&  e^2  M^{3-\epsilon}~ \mbox{Tr} \int_q 
\frac{(\gamma_i+2q_i+r_i)(\gamma_0(q_0+ r_0) -\gamma_l(q_l +r_l)+ (q+r)^2)}{\den ((q_0+r_0)^2 -(q+r)^2 -(q+r)^4)}\nn
&&~~~~~~~~~~~~~~~~~~~~~~~~~~\times(\gamma_j+2q_j+r_j)\num~.
\eea
As expected, given the configuration of the external space momentum, the off-diagonal terms and $\Pi_{33}$ vanish upon integration. 
We have 
\be
\Pi_{11}=\Pi_{22}= -Z
\ee
we find
\bea
Z&=& \frac{e^2}{M}  \frac{2^{\epsilon -5} \pi ^{\frac{\epsilon -3}{2}-\frac{1}{2}} (\epsilon -5) (\epsilon  (\epsilon +10)-3) 
\Gamma \left(-\frac{\epsilon }{2}\right)
   \Gamma \left(\frac{\epsilon +1}{2}\right)}{3 \Gamma \left(\frac{7}{2}-\frac{\epsilon }{2}\right)} \nonumber \\
&=&\frac{-e^2}{6\pi^2M} \left( \frac{-1}{\epsilon} +2 + \frac{A}{2} \right) +O(\epsilon)~.
\eea
The corresponding correction is then
\be
W-Z=\frac{11e^2}{18\pi^2M}~.
\ee

\end{document}